\documentstyle[12pt]{article}

\advance\hoffset by -0.5cm
\def\3{\ss}
\def\sq{\hbox{\rlap{$\sqcap$}$\sqcup$}}
\def\qed{\ifmmode\sq\else{\unskip\nobreak\hfil
\penalty50\hskip1em\null\nobreak\hfil\sq
\parfillskip=0pt\finalhyphendemerits=0\endgraf}\fi}
\def\bbbz {{\sf Z\!\!Z}}

\def\bbbone {{\mathchoice {\rm 1\mskip-4mu l} {\rm 1\mskip-4mu l}
{\rm 1\mskip-4.5mu l} {\rm 1\mskip-5mu l}}}
\def\bbbc{{\mathchoice {\setbox0=\hbox{$\displaystyle\rm C$}\hbox{\hbox
to0pt{\kern0.4\wd0\vrule height0.9\ht0\hss}\box0}}
{\setbox0=\hbox{$\textstyle\rm C$}\hbox{\hbox
to0pt{\kern0.4\wd0\vrule height0.9\ht0\hss}\box0}}
{\setbox0=\hbox{$\scriptstyle\rm C$}\hbox{\hbox
to0pt{\kern0.4\wd0\vrule height0.9\ht0\hss}\box0}}
{\setbox0=\hbox{$\scriptscriptstyle\rm C$}\hbox{\hbox
to0pt{\kern0.4\wd0\vrule height0.9\ht0\hss}\box0}}}}

\title{Fusion in conformal field theory as the tensor product of
the symmetry algebra}
\author{Matthias Gaberdiel
\thanks{e-mail: mg10004@amtp.cam.ac.uk} \\
Department of Applied Mathematics and Theoretical
Physics\\
University of Cambridge, Silver Street \\
Cambridge, CB3 9EW, U.\ K.\ }
\date{July 1993}

\begin{document}

\maketitle

\begin{abstract}
Following a recent proposal of Richard Borcherds
to regard fusion as the ring-like tensor product of modules
of a {\em quantum ring}, a generalization of rings and vertex
operators, we define fusion as a certain quotient of the (vector space)
tensor product of representations of the symmetry algebra ${\cal A}$.
We prove that this tensor product is associative and
symmetric up to equivalence. We also determine
explicitly the action of ${\cal A}$ on it, under
which the central extension is preserved.

Having given a precise meaning to fusion, determining
the fusion rules is now a well-posed algebraic problem,
namely to decompose the tensor product into
irreducible representations.
We demonstrate how to solve it for the case
of the WZW- and the minimal models
and recover thereby the well-known fusion rules.
\smallskip

The action of the symmetry algebra on the tensor product
is given in terms of a comultiplication. We calculate
the $R$-matrix of this comultiplication and find that it is
triangular. This seems to shed some new light
on the possible r\^{o}le of the quantum group in conformal
field theory.
\end{abstract}

\section{Introduction}

The notion of fusion plays a very important r\^{o}le in conformal
field theory, however, as far as the definition of fusion is concerned
the state of affairs is not entirely satisfactory.
There exist on the one hand the physicists' definition of fusion in terms
of nonvanishing three-point-functions, which has the advantage
of being easily calculable,  which however is conceptually not very
appealing. On the other hand there exist different definitions
in the mathematical literature, e.\ g.\ in terms of holomorphic embeddings
\cite{Segal},  more algebraically by Feigin and Fuchs \cite{FF}
(for the minimal models) and by Kazhdan and Lusztig \cite{KL} (for
the Kac-Moody algebras),
or in the spirit of algebraic quantum field theory \cite{Wasser},
however, these definitions tend to be rather abstract and it seems
quite difficult to perform any calculations in these setups.
It seems therefore quite interesting to give a precise mathematical
definition of fusion, which is at the same time sufficiently explicit
to allow for example a simple derivation of the fusion rules.
\smallskip

The present paper was motivated and inspired by the recent
proposal of Richard Borcherds \cite{B} to introduce the
concept of a {\em quantum ring}, a generalization of rings and
vertex algebras for which the holomorphic fields
are a natural example, to give a precise meaning to fusion.
{}From this point of view conformal field theory is just
the representation theory of this quantum ring and fusion
corresponds to taking the canonical ring-like tensor product
of modules, namely the
usual vector space tensor product quotiented by the relations which
guarantee that the action of the quantum ring on both
modules is the same. Richard Borcherds showed that
the corresponding tensor category is isomorphic to the
tensor category of Kazhdan and Lusztig, which is
believed to coincide with the physicists' definition of fusion.

To express this approach to conformal field theory
in more physical terms we have to relate the quantum ring of
holomorphic fields to the symmetry algebra, i.\ e.\
to the mode expansion of the holomorphic fields.
We use the methods of \cite{Peter89} to analyze conformal
field theory in detail and derive explicitly the
action of the symmetry algebra on products
of vertex operators. This action can
be interpreted as a comultiplication of the
symmetry algebra (depending on two
parameters, namely the insertion points of the
vertex operators), under which the central charge
is preserved, and can thus be used to define
a (vector space) tensor product of modules of the
symmetry algebra. (This comultiplication is a
generalization of the formula given in \cite{MS}.)

To derive the comultiplication formula we have to use
the operator product expansion of the holomorphic field
with one of the vertex operators. The formula we obtain is
thus quite asymmetric, as we have treated the two
vertex operators differently. If we used the operator
product expansion with the other vertex operator, we
would get a different formula, which, on the underlying conformal
field theory, must agree with the previous expression.
We cannot impose this equality on the level of the algebra, as
the two formulae are only formally well-defined
power series, however, we can implement it on the tensor product of
representations. This leads us to define
the ``true'' tensor product of conformal field theory to
be the quotient of the original (vector space)
tensor product by all relations which arise in this
way. By quotienting out this subspace
we guarantee that the action of the
holomorphic fields on both vertex operators is the same;
our tensor product corresponds thus exactly to
the ring-like tensor product of quantum ring modules.

The next step is to prove general properties of this
tensor product. Firstly, we determine the dependence
of the tensor product on the
two parameters and show that all
different choices are equivalent. We prove that the
comultiplication is coassociative and thus that the
tensor product is associative up to equivalence.
We then calculate the $R$-matrix of the comultiplication.
It turns out that it is triangular, which implies
that the comultiplication is not braided. To understand
the way in which the braiding appears in conformal field theory
we analyze the decomposition of the tensor product into irreducible
components. As the action of the symmetry algebra on the
tensor product depends on two parameters, an intertwiner mapping
the tensor product into an irreducible representation
will in general have some dependence on these parameters
as well. It turns out, that if one requires this intertwiner
to transform naturally under scaling and translation, it will
in general not be single valued, but exhibit the usual
braiding. This seems to indicate that the braiding
is no direct property of the symmetry algebra, but only
arises once one tries to decompose the tensor product
in a consistent way. In particular, we believe that
this indicates, that the quantum group relevant for
conformal field theory does not ``sit inside''
the symmetry algebra.

Finally, in section 4, we use the general framework to
find all the restrictions for the fusion rules of the WZW- and
the minimal models. Our method follows
in spirit  \cite{GW} and \cite{BPZ}, but is quite
different in flavour, as  it is
completely algebraic. In particular, having
proved general properties of the tensor product, we
can give a precise meaning to the intuitive argument of
\cite{BPZ}, which enables us to derive the general
restrictions for the fusion rules of
the minimal models from one special case.
To prove that these conditions are also sufficient,
we indicate how to show --- using the methods of
\cite{BFIZ}, \cite{FF}, \cite{K} ---
that the given representations are actually contained
in the tensor product.
This establishes that our definition of fusion
agrees with the usual one.

\section{Comultiplication in CFT}
\renewcommand{\theequation}{2.\arabic{equation}}
\setcounter{equation}{0}

By analyzing conformal field theory we derive the comultiplication  of
the symmetry algebra, which gives rise to the
vector space tensor product.
We demonstrate the method explicitly for the WZW-models and
give the result for the minimal models. This method can
easily be generalized to  other conformal field theories,
since we only use the $su(1,1)$-properties of the
chiral algebra. (We shall pursue this in a forthcoming paper
\cite{G}.)

\medskip

One way to think about fusion is that it keeps track of the
conformal families which occur in the operator product
expansion of two vertex operators. To obtain the operator
product expansion of two vertex operators one has to bring --- at
least heuristically ---  the two operators close
together and determine the action of the symmetry algebra
in the limit, where the two points coincide. In general
this limit is singular and the expansion in the difference
of the two insertion points contains fractional powers,
reflecting the braid group statistics of the vertex operators.
To avoid  performing this limit one may alternatively consider
the limit in which the holomorphic field (whose action on
the fused state we want to determine) tends to infinity for
fixed insertion points of the two vertex operators. Then
one has to calculate the contour integral (over a large contour $C$,
i.\ e.\
a contour encircling the two insertion points)
\begin{equation}
\label{cont}
 \oint_{C} dw \; w^{m} \;J^{a}(w) \;V(\psi,\zeta) V(\chi,z)\;\Omega ,
\end{equation}
where $V(\psi,\zeta)$ and $V(\chi,z)$ denote the two vertex
operators, $\Omega$ is the vacuum, $J^{a}(w)$  the current
field in the symmetry algebra and $|\zeta|>|z|$ in order to guarantee
the existence of the operator product.
Assuming that the vertex operator
and the current field are local with respect to each other
one can apply the methods of \cite{Peter89} to obtain the
operator product expansion of the two fields
\begin{equation}
\label{pg}
J^{a}(w)\;V(\psi,\zeta) = \sum_{l\in\bbbz} (w-\zeta)^{l-1}\;
V(J^{a}_{-l}\psi,\zeta).
\end{equation}
(If $\psi$ belongs to the dense subspace of finite energy
states, the right-hand-side contains only a pole of finite order.)
Then one can use the usual contour deformation arguments to evaluate
$(\ref{cont})$, as the integrand has only poles at $0,\; z$ and $\zeta$ and
$(\ref{pg})$ determines the residues. However, if we did the calculation
at this stage, we would get (for $m<0$) a contribution from $w=0$
of the form
\begin{equation}
V(\psi,\zeta)\;V(\chi,z)\;J^{a}_{m}\;\Omega,
\end{equation}
and this term would not allow an interpretation as a comultiplication.
We can circumvent this problem in the following way. Considering the
scalar product of the integrand of (\ref{cont}) with any vector
$\varphi\in{\cal F}$, the dense subspace of ${\cal H}$ of finite
energy vectors
\begin{equation}
\left\langle\varphi, J^{a}(w) \;V(\psi,\zeta)\; V(\chi,z)\;\Omega\right\rangle
\end{equation}
we get a meromorphic function of $w$, whose poles are determined
by (\ref{pg}). Thus subtracting these poles we get an entire function,
namely
$$
\left\langle\varphi, J^{a}(w) \;V(\psi,\zeta)\;
V(\chi,z)\;\Omega\right\rangle -
\sum_{m=-\infty}^{0} (w-\zeta)^{m-1} \langle\varphi,V(J^{a}_{-m}\psi,\zeta)\;
V(\chi,z)\;\Omega\rangle \hspace*{0.5cm}
$$
\begin{equation}
\label{poles}
\hspace*{3.5cm}
-\sum_{l=-\infty}^{0} (w-z)^{l-1}
\langle\varphi, V(\psi,\zeta) \; V(J^{a}_{-l}\chi,z)\;\Omega\rangle.
\end{equation}
Using the operator product expansion of the holomorphic
field $J^{a}(w)$ with the vertex operator at $z$ we can write
this function as
\begin{equation}
\label{exp}
\sum_{l=1}^{\infty} (w-z)^{l-1} \langle\varphi,V(\psi,\zeta)\;
V(J^{a}_{-l}\chi,z)\;\Omega\rangle -
\sum_{m=-\infty}^{0} (w-\zeta)^{m-1}
\langle\varphi, V(J^{a}_{-m}\psi,\zeta) \; V(\chi,z)\;\Omega\rangle.
\end{equation}
Rewriting (\ref{exp}) as a power series about $w=0$ (assuming
that it makes sense for $w=0$, i.\ e.\ that  $\zeta$
and $z$ are suitably chosen) the series will converge for
all $w$, as (\ref{exp}) is entire.
As the integrand in (\ref{cont}) is just this entire function
plus the poles we have subtracted in (\ref{poles}), we can now
use this power series to evaluate the contour integral,
thereby obtaining
only terms where $J^{a}_{m}$ either acts on $\psi$ or on $\chi$.
We remark that the action of $J$ on the operator product is
independent of the vector $\varphi$. We can thus interpret
the formula as giving a comultiplication of the Kac-Moody algebra,
namely
$$
\oint_{C} dw \; w^{m}
\left\langle\varphi, J^{a}(w) \;V(\psi,\zeta) \;
V(\chi,z)\;\Omega \right\rangle
= \hspace*{3.5cm}
$$
\begin{equation}
\label{m}
\hspace*{3.5cm} \sum
\left\langle\varphi, V(\Delta^{(1)}_{\zeta,z}\,(J^{a}_{m})\,\psi,\zeta)\;
V(\Delta^{(2)}_{\zeta,z}\,(J^{a}_{m})\,\chi,z)\;\Omega \right\rangle ,
\end{equation}
where we write $\Delta_{\zeta,z}(a)\in{\cal A}\otimes{\cal A},\;
a\in{\cal A}$ as
\begin{equation}
\Delta_{\zeta,z}(a) = \sum \Delta_{\zeta,z}^{(1)}(a)\otimes
\Delta_{\zeta,z}^{(2)}(a).
\end{equation}
We calculate
\begin{eqnarray}
\label{wzw1}
{\displaystyle \Delta_{\zeta,z}(J^{a}_{n})} = &
{\displaystyle \sum_{m=0}^{n} \left( \begin{array}{c} n \\ m
\end{array} \right)
\zeta^{n-m} \left(J^{a}_{m} \otimes \bbbone\right) +
\sum_{l=0}^{n} \left( \begin{array}{c} n \\ l
\end{array} \right)
z^{n-l} \left(\bbbone \otimes J^{a}_{l} \right)} \hspace*{2.5cm} \\
{\displaystyle \Delta_{\zeta,z}(J^{a}_{-n})} = &
{\displaystyle \sum_{m=0}^{\infty} \left( \begin{array}{c} n+m-1 \\ m
\end{array} \right) (-1)^{m}
\zeta^{-(n+m)} \left(J^{a}_{m} \otimes \bbbone \right) }
\hspace*{3.7cm} \nonumber \\
\label{wzw2}
& \hspace*{3.5cm} {\displaystyle
+ \sum_{l=n}^{\infty} \left( \begin{array}{c} l-1 \\ n-1
\end{array} \right)
(-z)^{l-n} \left(\bbbone \otimes J^{a}_{-l}  \right)},
\end{eqnarray}
where in (\ref{wzw1}) $n\geq 0$ and in (\ref{wzw2}) $n\geq 1$.
\medskip

Similarly, we can use the operator product expansion of the
energy-momentum tensor with a vertex operator, e.\ g.\ as given in
\cite{Peter89}
\begin{equation}
T(w)\;V(\psi,\zeta)=\sum_{l\in\bbbz}\;(w-\zeta)^{l-2}\;V(L_{-l}
\psi,\zeta)
\end{equation}
to derive the comultiplication for the Virasoro algebra, thereby
obtaining
\begin{equation}
\label{vir1}
{\displaystyle \Delta_{\zeta,z}(L_{n})} =
{\displaystyle \sum_{m=-1}^{n} \left( \begin{array}{c} n+1 \\ m+1
\end{array} \right)
\zeta^{n-m} \left(L_{m} \otimes \bbbone \right) +
\sum_{l=-1}^{n} \left( \begin{array}{c} n+1 \\ l+1
\end{array} \right)
z^{n-l} \left(\bbbone \otimes L_{l} \right)} \hspace*{0.4cm}
\end{equation}
\begin{eqnarray}
{\displaystyle \Delta_{\zeta,z}(L_{-n})} = &
{\displaystyle \sum_{m=-1}^{\infty} \left( \begin{array}{c} n+m-1 \\ m+1
\end{array} \right) (-1)^{m+1}
\zeta^{-(n+m)} \left(L_{m} \otimes \bbbone \right) } \hspace*{2.5cm}
\nonumber \\
\label{vir2}
& \hspace*{4.5cm} {\displaystyle
+ \sum_{l=n}^{\infty} \left( \begin{array}{c} l-2 \\ n-2
\end{array} \right)
(-z)^{l-n} \left(\bbbone \otimes L_{-l} \right)},
\end{eqnarray}
where in (\ref{vir1}) $n\geq -1$ and in (\ref{vir2}) $n\geq 2$.
These formulae are generalizations of the comultiplication formulae
given in \cite{MS}, to which they reduce if we set $z=0$.
\smallskip

Formally the sums converge for $|\zeta|>1$ and $|z|<1$.
If we restrict ourselves to a suitable dense subset
of the tensor product (containing all tensor products
of finite energy vectors) the comultiplication is
well-defined for (a suitable subset of) $|z|<1$ and
$\zeta\in\bbbc\backslash\{ 0 \}$.

The representation induced by the comultiplication is
a highest weight representation in the sense that
for given highest weight vectors $\psi$ and $\chi$
\begin{equation}
\label{hwv}
\Psi:= e^{-\zeta L_{-1}} \psi \otimes e^{-z L_{-1}} \chi
\end{equation}
is a highest weight vector for the action of
$\Delta_{\zeta,z}$, i.\ e.\
$\Delta_{\zeta,z}\,(J^{a}_{n})\; \Psi=0$ for $n>0$ and
similarly for the Virasoro case.
(This follows from the proof of (\ref{para}) below.)
However, $\Psi$ does not belong to the domain of definition for the
comultiplication, as the action of
$\Delta_{\zeta,z}\,(J^{a}_{-n})$ is not well-defined
on $\Psi$. We shall not discuss these problems in this paper.
\medskip

To prove the comultiplication-property of these formulae, we
have to check, that
\begin{equation}
\label{com}
[ \Delta_{\zeta.z} (a), \Delta_{\zeta.z} (b) ] =   \Delta_{\zeta.z}
([a,b]) \hspace{0.5cm} \mbox{for all $a,b \in {\cal A}$},
\end{equation}
where ${\cal A}$ is the symmetry algebra, i.\ e.\ the Kac-Moody-
or the Virasoro algebra, respectively.
As this is quite technical we have banned an outline of the proof
into appendix A.
\medskip

The comultiplication formulae can be thought of as being the
mode expansion of operator products involving a vertex operator
and a holomorphic field. From this point of view
the comultiplication-property  reflects
the fact, that the subalgebra of holomorphic fields
is closed (and associative). However, as we derive these
formulae form a well-defined conformal field theory, they
will satisfy further relations reflecting
various other properties of the theory. Some of these
properties will be automatically satisfied on the level
of the algebra (e.\ g.\ the comultiplication-property),
however, as we shall see in the following, there are also
relations, which do not hold true on the level of the {\em algebra}
and which have to be implemented on the level of {\em representations}
in order to maintain the whole information of conformal
field theory.

\section{The ring-like tensor product}
\renewcommand{\theequation}{3.\arabic{equation}}
\setcounter{equation}{0}

The comultiplication formulae
(2.9-10,\, 2.12-13)
we have derived from conformal field theory are asymmetric in the
sense that they treat the two algebras in the tensor product on
a different footing. This is quite unsatisfactory, as one expects
on general grounds that the symmetry algebra possesses a
quasitriangular-like structure, reflecting the locality
of the theory.

To understand how to resolve this problem, let us recall, how
we derived the comultiplication. (In the following
we restrict ourselves to the case of the WZW-models; obviously
similar statements hold for the minimal models.)
Assuming that $|\zeta|>|z|$
in order to guarantee the existence of the operator product, we
used the operator product expansion of the holomorphic field at $w$ with the
vertex operator at $z$ to get an expansion for the entire
function (\ref{poles}), which could then be integrated. The
comultiplication we obtained was well-defined for $|z|<1$ and
$\zeta\in\bbbc\backslash \{ 0 \}$, if we restricted ourselves
to the action on tensor products of finite energy vectors.
However, we could have
carried through the whole program with the r\^{o}les  of $z$ and
$\zeta$ interchanged, thereby getting instead of (\ref{wzw2})
\begin{eqnarray}
\label{wzw2'}
{\displaystyle {\widetilde \Delta}_{\zeta,z}(J^{a}_{-n})} & = &
{\displaystyle \sum_{m=n}^{\infty} \left( \begin{array}{c} m-1 \\ n-1
\end{array} \right)
(-\zeta)^{m-n} \left(J^{a}_{-m} \otimes \bbbone \right) }
\hspace*{3.5cm} \nonumber \\
& & \hspace*{3.0cm} {\displaystyle
+ \sum_{l=0}^{\infty} \left( \begin{array}{c} n+l-1 \\ l
\end{array} \right) (-1)^{l}
z^{-(n+l)} \left(\bbbone \otimes J^{a}_{l}  \right)}.
\end{eqnarray}
Again this formula is well-defined for $|\zeta|<1$ and
$z\in\bbbc\backslash\{ 0 \}$ if we restrict ourselves to the action
on tensor products of finite energy vectors.
Thus in particular, for
$z,\zeta \in \{u\in\bbbc |\;  |u|<1, u\neq 0 \}$
the two formulae (\ref{wzw2}) and (\ref{wzw2'}) should agree
when restricted to this subset of vectors. (In fact, as we can
see from (A.8) and (A.9), the two formulae are just
different power series expansions
of one analytic function.)
However, this relation is not true on the level of algebras, and
in order to maintain the whole information of conformal field theory
we have to {\em impose} it on the level of representations.
This is most naturally done in a way which closely resembles
Richard Borcherds  \cite{B} proposal to regard fusion in conformal
field theory as a ring-like tensor product of modules of
the ``quantum ring'' of (holomorphic) fields. Namely,
instead of defining the tensor product to be the usual (vector space)
tensor product of representations, we define it to be the
quotient of this vector space by all relations
guaranteeing the equality of (\ref{wzw2}) and (\ref{wzw2'}).
This space is then the ``true'' tensor product of conformal field
theory. The action of the symmetry algebra on it is given
by either of the two comultiplication formulae.
\bigskip

The next step is to determine various general properties
of this tensor product.
Firstly we observe, that for  $|u|<|\zeta|$  we have
\begin{equation}
\label{para}
(e^{u L_{-1}}\otimes e^{v L_{-1}})\circ\Delta_{\zeta +u,z +v}\circ
(e^{-u L_{-1}}\otimes e^{-v L_{-1}}) = \Delta_{\zeta,z}
\end{equation}
and similarly for $|v|<|z|$
\begin{equation}
\label{para'}
(e^{u L_{-1}}\otimes e^{v L_{-1}})\circ
{\widetilde \Delta}_{\zeta +u,z +v}\circ
(e^{-u L_{-1}}\otimes e^{-v L_{-1}}) = {\widetilde \Delta}_{\zeta,z}.
\end{equation}
The proof of (\ref{para}), resp. (\ref{para'})  is just another
standard manipulation of sums, once one observes that
\begin{equation}
\label{lmw}
e^{u L_{-1}}\; J^{a}_{m} \; e^{-u L_{-1}} = \left\{
\begin{array}{ll}
{\displaystyle \sum_{l=0}^{m}
\left( \begin{array}{c} m \\ l \end{array} \right)
(-u)^{m-l} J^{a}_{l}} & \mbox{if $m\geq 0$} \\
{\displaystyle \sum_{l=-m}^{\infty}
\left( \begin{array}{c} l-1 \\ -m-1 \end{array} \right)
u^{l+m} J^{a}_{-l}} & \mbox{if $m \leq -1$},
\end{array}
\right.
\end{equation}
for $u$ sufficiently small.
\smallskip

\noindent (\ref{para}) and (\ref{para'}) imply that the tensor
products corresponding to different values of the
parameters are equivalent, as we can define an
intertwiner $U$, well-defined on the quotient
$$
U: {\cal H}_{1} \otimes_{\zeta,z} {\cal H}_{2}
\rightarrow {\cal H}_{1} \otimes_{\zeta+u,z+v} {\cal H}_{2}
$$
\begin{equation}
(\psi\otimes_{\zeta,z} \chi) \mapsto (e^{-u L_{-1}} \psi
\otimes_{\zeta+u,z+v} e^{-v L_{-1}} \chi),
\end{equation}
such that
\begin{equation}
\begin{array}{rcl}
{\displaystyle
\Delta_{\zeta,z}}
&  = &
{\displaystyle U^{-1}\circ\Delta_{\zeta+u,z+v}\circ U }\\
{\displaystyle
{\widetilde \Delta}_{\zeta,z}}
& =  &
{\displaystyle U^{-1}\circ
{\widetilde \Delta}_{\zeta+u,z+v}\circ U.}
\end{array}
\end{equation}
Strictly speaking this is only true for those values of
$\zeta, z, u$ and $v$ for which (\ref{para}) and (\ref{para'}) make
sense. However, as we may apply to the conformal nature
of the theory, i.\ e.\ that the comultiplication
must depend analytically on the parameters $\zeta$ and $z$,
it is for example sufficient to prove the fusion rules
for open neighbourhoods of suitably chosen values and
thus, because of the above, only for one pair
$(\zeta,z)$. (We will use this argument in the derivation
of the fusion rules of the minimal models.)
\medskip

As the operator product of the underlying conformal field theory
is associative we expect the comultiplication to be
coassociative. In fact, for suitable $\zeta_{1}, \zeta_{2}, z$
and $w$ we have
\begin{equation}
\label{coass}
\begin{array}{rcl}
{\displaystyle
\left( \Delta_{\zeta_{2} - w,\zeta_{1}-w}\otimes\bbbone \right)
\circ \Delta_{w,z}} & = &
{\displaystyle
\left( \bbbone\otimes \Delta_{\zeta_{1}-w,z-w} \right) \circ
\Delta_{\zeta_{2},w}} \\
{\displaystyle
\left( \Delta_{\zeta_{2} - w,\zeta_{1}-w}\otimes\bbbone \right)
\circ \widetilde{\Delta}_{w,z}} & = &
{\displaystyle
\left( \bbbone\otimes \widetilde{\Delta}_{\zeta_{1}-w,z-w} \right) \circ
\Delta_{\zeta_{2},w}} \\
{\displaystyle
\left( \widetilde{\Delta}_{\zeta_{2} - w,\zeta_{1}-w}
\otimes\bbbone \right)
\circ \widetilde{\Delta}_{w,z}} & = &
{\displaystyle
\left( \bbbone\otimes \Delta_{\zeta_{1}-w,z-w} \right) \circ
\widetilde{\Delta}_{\zeta_{2},w}} \\
{\displaystyle
\left( \widetilde{\Delta}_{\zeta_{2} - w,\zeta_{1}-w}
\otimes\bbbone \right)
\circ \Delta_{w,z}} & = &
{\displaystyle
\left( \bbbone\otimes \Delta_{\zeta_{1}-w,z-w} \right) \circ
\Delta_{\zeta_{2},w}} \\
{\displaystyle\left( \widetilde{\Delta}_{\zeta_{2} -w,\zeta_{1}-w}
\otimes\bbbone \right)
\circ \widetilde{\Delta}_{w,z}} & = &
{\displaystyle
\left( \bbbone\otimes \widetilde{\Delta}_{\zeta_{1}-w,z-w} \right) \circ
\widetilde{\Delta}_{\zeta_{2},w},}
\end{array}
\end{equation}
where $\widetilde{\Delta}$ denotes the comultiplication
(\ref{wzw1},\,\ref{wzw2'}) and $\Delta$ the original one
(\ref{wzw1},\,\ref{wzw2}). The proof of these equations is
again straight forward; for the convenience of the reader
we have included the proof of one special case
in appendix B.
\smallskip

The above equations imply that the triple tensor
product of conformal field theory is well-defined,
as the subspace, by which we have to quotient the
vector space tensor product, is independent of
the bracketing. In addition, the action of the
symmetry algebra on this tensor product is
independent of the bracketing if we choose the
parameters according to (\ref{coass}). Thus the
tensor product is associative up to equivalence,
as tensor products corresponding to different
values of the parameters are equivalent.
\smallskip

We remark that using the same methods as in section 2
we could have also determined the action
of the symmetry algebra on products of three vertex operators.
A straight forward calculation shows
that the action (\ref{coass}) agrees with this action if the
vertex operators are inserted at $\zeta_{2}, \zeta_{1}$ and $z$.
(The different expressions in (\ref{coass})
correspond to which operator product expansion we choose
in (\ref{exp}).)
Thus fusion of $n$ representations really corresponds to taking
successively $(n-1)$ tensor products.
\medskip

\noindent Thirdly, the $L_{0}$ action of conformal field
theory translates into
\begin{equation}
\label{scal}
\begin{array}{rcl}
{\displaystyle
\left(\lambda^{L_{0}} \otimes \lambda^{L_{0}} \right)\;
\Delta_{\zeta,z}(A)\;
\left(\lambda^{-L_{0}} \otimes \lambda^{-L_{0}} \right)}
& = &
{\displaystyle
 \Delta_{\lambda\zeta, \lambda z}\left(\lambda^{L_{0}} A
\lambda^{-L_{0}}\right)} \\
{\displaystyle
\left(\lambda^{L_{0}} \otimes \lambda^{L_{0}} \right)\;
{\widetilde \Delta}_{\zeta,z}(A)\;
\left(\lambda^{-L_{0}} \otimes \lambda^{-L_{0}} \right)}
& = &
{\displaystyle
{\widetilde{\Delta}_{\lambda\zeta, \lambda z}\left(\lambda^{L_{0}} A
\lambda^{-L_{0}}\right)},}
\end{array}
\end{equation}
which can be checked very easily.
\medskip

Finally, using the fact, that the two different comultiplications agree
on these tensor products, we can find the universal $R$-matrix
of the comultiplication,
i.\ e.\ the operator $R(\zeta,z)$ in ${\cal A}\otimes {\cal A}$,
which satisfies on all tensor products
\begin{equation}
\label{rm}
R(\zeta,z)\circ\Delta_{\zeta,z}\circ R(\zeta,z)^{-1} =
\tau\circ\Delta_{\zeta,z},
\end{equation}
where $\tau$ is the
twist-map, interchanging the two factors of the tensor product.
In fact, using (\ref{para}), one can easily show that $R(\zeta,z)$ is given by
\begin{equation}
\label{R}
R(\zeta,z)=e^{(\zeta - z) L_{-1}} \otimes e^{(z - \zeta) L_{-1}},
\end{equation}
where we assume, that $\zeta$ and $z$ are suitably chosen to ensure
convergence. At first sight (\ref{R}) might seem quite surprising,
as the $R$-matrix is triangular, i.\ e.\
\begin{equation}
\label{RR}
\tau\left(R(\zeta,z)^{-1}\right)=R(\zeta,z),
\end{equation}
which implies, that the comultiplication is not braided.
However, recalling the definition of the comultiplication
from conformal field theory (\ref{m}), we can easily understand, why
this has to be true: $\Delta^{(i)}_{\zeta,z}(J^{a}_{m})$ is just
a Kac-Moody generator (or $\bbbone$) and the monodromy
properties of the scalar product depend only on the
conformal families, hence the monodromy of the
right and left hand side can only be equal, if $\Delta_{\zeta,z}$ has
trivial monodromy, which is precisely (\ref{RR}).
(Another way of understanding (\ref{RR}) is, that it
reflects the fact that the action of the holomorphic fields
on the vertex operators is local.)
We would like to point out that this does {\em not} imply that
the corresponding tensor category is not braided. For our
comultiplication is not braided,  but also not truly
coassociative, as we have (\ref{coass})
$$
\left( e^{\zeta_{1} L_{-1}} \otimes \bbbone \otimes \bbbone \right)
\left( (\Delta_{\zeta_{2},\zeta_{1}} \otimes \bbbone) \circ
\Delta_{\zeta_{1},z} \right)
\left( e^{-\zeta_{1} L_{-1}} \otimes \bbbone \otimes \bbbone \right)
\hspace*{4.5cm}
$$
\begin{equation}
\hspace*{3.5cm}
=\left(\bbbone\otimes\bbbone\otimes e^{\zeta_{1} L_{-1}} \right)
\left( (\bbbone\otimes \Delta_{\zeta_{1},z}) \circ
\Delta_{\zeta_{2},\zeta_{1}} \right)
\left(\bbbone\otimes\bbbone\otimes e^{-\zeta_{1} L_{-1}} \right).
\end{equation}
(In fact, choosing all parameters to be the same, i.\ e.\
$\zeta=z\neq 0$, the comultiplication is symmetric and
quasi-coassociative.)
\medskip

To obtain some better understanding of the r\^{o}le of the braiding
in conformal field theory let us next analyze the decomposition of the
tensor product into irreducible components.
Suppose the representation ${\cal H}_{3}$ is contained in
the tensor product of two representations ${\cal H}_{i}\otimes_{\zeta,z}
{\cal H}_{2}$, where the action of the symmetry algebra ${\cal A}$
is given by $\Delta_{\zeta,z}$. Then there exists an intertwiner
for the action of the symmetry algebra
\begin{equation}
\label{int}
\pi_{\zeta,z}: {\cal H}_{1} \otimes_{\zeta,z} {\cal H}_{2} \rightarrow
    {\cal H}_{3},
\end{equation}
i.\ e.\ a vector space homomorphism satisfying
\begin{equation}
\pi_{\zeta,z} \circ \Delta_{\zeta,z}(a) = a \circ \pi_{\zeta,z}
\end{equation}
for all $a\in {\cal A}$, and a vector $\psi_{3}\in{\cal H}_{3}$, such that
the scalar product
\begin{equation}
F(\zeta,z)=\left\langle\psi_{3},\pi_{\zeta,z}\left(\psi_{1}
\otimes_{\zeta,z} \psi_{2}\right)
\right\rangle
\end{equation}
does not vanish for all $\psi_{i}\in{\cal H}_{i},\;i=1,2$.
By Schur's Lemma, the intertwiner $\pi_{\zeta.z}$ is only determined
up to a scalar (assuming no multiplicities, for simplicity).
However, because of (\ref{para}) and (\ref{scal}) there is
a natural choice to relate the intertwiners for different
values of $\zeta$ and $z$, as we can choose
\begin{equation}
\pi_{\zeta -u, z-u} = e^{-u L_{-1}} \circ \pi_{\zeta,z}
\end{equation}
and
\begin{equation}
\pi_{\lambda\zeta,\lambda z} = \lambda^{L_{0}} \circ \pi_{\zeta,z}\circ
\left(\lambda^{-L_{0}} \otimes \lambda^{-L_{0}}\right),
\end{equation}
where in the second equation it is understood, that a choice for the
logarithm of $\lambda$ has been made. This fixes the intertwiner
up to an overall phase factor, if we allow for multivaluedness.
If $\psi_{3}\in{\cal H}_{3}$ is now a Virasoro highest weight
vector, then the above two equations imply that
\begin{equation}
\label{drei}
F(\zeta,z)= C(\psi_{1},\psi_{2},\psi_{3})\;
(\zeta-z)^{\Delta_{3}-\Delta_{1}-\Delta_{2}},
\end{equation}
where $L_{0}\psi_{i}=\Delta_{i} \psi_{i},\;i=1,2,3$.
Thus the scalar product, which corresponds to the
three point function of conformal field theory,
exhibits the expected braiding. However, the braiding
does not seem to be a direct property of the
symmetry algebra; it rather appears as we try to decompose
the tensor product in a consistent way.

\section{Fusion rules}
\renewcommand{\theequation}{4.\arabic{equation}}
\setcounter{equation}{0}

In this section we want to show, how to derive the fusion rules
of the WZW- and  the  minimal models from the above
definition of fusion. We derive the restrictions for
the possible fusions following  essentially \cite{GW} and \cite{BPZ},
but using an entirely algebraic language. To prove that
these restrictions are sufficient in the case of the minimal
models, we refer to the proofs of \cite{BFIZ}, \cite{FF},
which can be easily adapted to the present situation.

Let us first consider the WZW-models, where the symmetry
algebra is the affine Lie algebra
$\hat{g}$ corresponding to the Lie algebra $g$ at level $x=2k/\psi^{2}$
($\psi$ is any long root of $g$ and $k$ is the central parameter).
The irreducible unitary positive energy representations of this
affine Lie algebra for fixed level $x$ are labeled by the
highest weight representations of the Lie algebra $g$, where in order
for a representation of $g$ to be a highest weight representation
of $\hat{g}$ the corresponding weight $\lambda$ must satisfy
$|\langle\lambda,\beta\rangle| \leq k$ for all roots $\beta$
\cite{GW}, \cite{Peter88}.

The Kac-Moody representation ${\cal H}_{3}$ is contained
in the tensor product
of two $\hat{g}$-re\-pre\-sen\-ta\-tions
${\cal H}_{1}\otimes_{\zeta,z} {\cal H}_{2}$,
if there exists an intertwiner for
the action of the Kac-Moody algebra
$\pi_{\zeta,z}: {\cal H}_{1} \otimes_{\zeta,z} {\cal H}_{2}
\rightarrow   {\cal H}_{3},$
and a vector $\phi$ in
the highest weight representation of ${\cal H}_{3}$, such that
the scalar product
$\left\langle\phi,\pi_{\zeta,z}\left(\psi_{1}\otimes_{\zeta,z}
\psi_{2}\right)
\right\rangle$
does not vanish for all highest weight vectors $\psi_{i},\;i=1,2$.
(In the limit $\zeta, z \rightarrow 0$ the right-hand sides
of the scalar products
become the highest weight vectors of the tensor product --- cf.\
the discussion following (\ref{hwv}). However,
we do not have to take this limit, as the scalar products
either vanish identically or are non-zero for all $\zeta\neq z$,
because of (\ref{drei}).)
We can immediately observe --- by considering the comultiplication
(\ref{wzw1}) for $n=0$ --- that a $\hat{g}$-representation will only
be contained in the $\hat{g}$-tensor product of two
$\hat{g}$-representations, if the corresponding highest weight
representation is contained in the $g$-tensor product of the
corresponding highest weight representations.
To obtain further restrictions we observe --- following
\cite{GW}, \cite{Peter88} --- that the three generators
\begin{equation}
J^{-\theta}_{+1} \hspace{0.5cm}
J^{\;\theta}_{-1} \hspace{0.5cm}
\frac{2}{\theta^{2}} \left( k - \theta \cdot H_{0} \right)
\end{equation}
satisfy the relations of an $su(2)$-subalgebra, where
$H_{0}$ denotes the elements of a Cartan subalgebra of $g$
and $J^{\theta}$ is a step operator corresponding to a root $\theta$
of $g$.
Now, suppose that $\phi\in{\cal H}_{3}$ is a highest weight vector
with weight $\phi$. Writing
$M=\frac{2}{\theta^{2}}(k - \phi\cdot\theta)$ the above implies
\begin{equation}
(J^{\theta}_{-1})^{M+1} \phi =0,
\end{equation}
as $J^{-\theta}_{+1} \phi=0$. Hence for all tensor products
of highest weight vectors
$\psi_{i}\in{\cal H}_{i},\; i=1,2$ we have (in the following
we suppress the subscripts of the tensor products)
\begin{eqnarray}
0 & = &
{\displaystyle
\left\langle (J^{\theta}_{-1})^{M+1}\phi,\pi_{\zeta,z}\left(\psi_{1}
\otimes\psi_{2}\right)\right\rangle} \nonumber \\
& = &
{\displaystyle
\left\langle  \phi,\pi_{\zeta,z}\left(\Delta_{\zeta,z}(J^{-\theta}_{+1})^{M+1}
\left(\psi_{1}\otimes\psi_{2}\right)\right)
\right\rangle} \nonumber \\
& = &
{\displaystyle
\sum_{l=0}^{M+1}
\left( \begin{array}{c} M+1 \\ l \end{array} \right)
\zeta^{l} z^{M+1-l}
\left\langle  \phi,\pi_{\zeta,z}\left(
(J^{-\theta}_{0})^{l}\psi_{1}\otimes (J^{-\theta}_{0})^{M+1-l}\psi_{2}\right)
\right\rangle.}
\end{eqnarray}
All summands have to vanish separately, as we can vary
$\zeta$ and $z$ independently and the $(\zeta,z)$ dependence of
all scalar products is the same, as a consequence of (\ref{drei}).
Writing $\phi_{1}=(J^{-\theta}_{0})^{l_{1}}\psi_{1}$ and
$\phi_{2}=(J^{-\theta}_{0})^{M+1-l}\psi_{2}$
we recover \cite[(3.49)]{GW}, from which one can derive
the ``depth rule'' of Gepner and Witten \cite{GW}.
\medskip

Next, we turn to deriving the fusion rules for the minimal
models. Following the notation of \cite{BPZ}
the primary fields $\psi_{(m,n)}$ of the theory are parametrized
by two positive integers $m$ and $n$, giving the conformal
weights of the corresponding fields via the equation
\begin{equation}
\Delta_{(m,n)}=\Delta_{0} + \left( \frac{1}{2} \alpha_{+} m +
\frac{1}{2} \alpha_{-} n \right)^{2},
\end{equation}
where
\begin{equation}
\Delta_{0}=\frac{1}{24} \left( c-1 \right),
\end{equation}
\begin{equation}
\alpha_{\pm}=\frac{\sqrt{1-c} \pm \sqrt{25-c}}{\sqrt{24}}
\end{equation}
and $c$ is the central charge of the Virasoro algebra, $0\leq c < 1$.
The primary fields of the theory are in one-to-one
correspondence with a certain subset of the irreducible, positive
energy representations of the Virasoro algebra  and we denote the
representation space corresponding to the field $\psi_{(m,n)}$ by
${\cal H}_{(m,n)}$. (The primary field $\psi_{(m,n)}$ is then a cyclic
vector in ${\cal H}_{(m,n)}$ for the action of the Virasoro
algebra.)

We restrict ourselves (without loss of generality) to
analyzing the decomposition of the tensor product
into irreducible representations for the case when
$z=0$ and $\zeta\neq 0$, as the representations for
different values of $z$ and $\zeta$ are equivalent, because of (\ref{para}).
Let us thus suppose that the representation ${\cal H}_{3}$
is contained in the tensor product
of the two representations ${\cal H}_{(m,n)}$
and ${\cal H}_{(1,2)}$ (resp. ${\cal H}_{(2,1)}$). Then
there exists an intertwiner (for the action of the Virasoro algebra)
$\pi: {\cal H}_{(m,n)} \otimes {\cal H}_{(1,2)} \rightarrow
{\cal H}_{(m',n')},$
such that the scalar product
$\left\langle\psi_{(m',n')},\pi\left(\psi_{(m,n)}\otimes\psi_{(1,2)}\right)
\right\rangle$
does not vanish.
Now, by definition, $\psi_{(1,2)}$ satisfies the
equation
\footnote{In formula (5.2) of \cite{BPZ} the coefficient of
$L_{-1}^{2}$ has the wrong sign.}
\begin{equation}
 \left(L_{-2} - \frac{3}{2(2\Delta_{(1,2)} +1)} L_{-1}^{2}\right)
 \psi_{(1,2)} = 0.
\end{equation}
As $\psi_{(m'n')}$ is a highest weight vector we thus obtain
\begin{eqnarray}
\label{null}
0 & = &
{\displaystyle \left\langle\psi_{(m',n')}, \pi\left( \left(
\Delta_{\zeta,0} (L_{-2}) - \frac{3}{2(2\Delta_{(1,2)} +1)}
\Delta_{\zeta,0} (L_{-1})^{2}\right)
 (\psi_{(m,n)}\otimes\psi_{(1,2)})\right) \right\rangle} \nonumber \\
& = &
{\displaystyle \left\langle\psi_{(m',n')}, \pi\left( \left( (\bbbone\otimes
L_{-2}) - \frac{3}{2(2\Delta_{(1,2)} +1)} (\bbbone\otimes L_{-1}^{2})\right)
 (\psi_{(m,n)}\otimes\psi_{(1,2)})\right) \right\rangle} \nonumber \\
& &
{\displaystyle + \left\langle\psi_{(m',n')}, \pi\left( \left(
\zeta^{-1} (L_{-1}\otimes\bbbone) - \zeta^{-2}
(L_{0}\otimes\bbbone)\right. \right. \right.}
\nonumber \\
& &
{\displaystyle - \left. \left. \left. \frac{3}
{2(2\Delta_{(1,2)} +1)} \left(2(L_{-1}\otimes L_{-1}) + (L_{-1}^{2}
\otimes\bbbone)\right) \right)  (\psi_{(m,n)}\otimes\psi_{(1,2)})\right)
\right\rangle} \nonumber \\
& = &
{\displaystyle \left\langle\psi_{(m',n')}, \pi\left( \left( \zeta^{-1} (
L_{-1}\otimes\bbbone) - \zeta^{-2} \Delta_{(m,n)} \right. \right. \right.}
\nonumber \\
& &
{\displaystyle - \left. \left. \left. \frac{3}
{2(2\Delta_{(1,2)} +1)} \left(2(L_{-1}\otimes L_{-1}) + (L_{-1}^{2}
\otimes\bbbone)\right) \right)  (\psi_{(m,n)}\otimes\psi_{(1,2)})\right)
\right\rangle.}
\end{eqnarray}
To calculate the terms containing $L_{-1}$ we insert $L_{0}$ into the
scalar product, using
$\Delta_{\zeta,0}(L_{0}) = (L_{0} \otimes \bbbone)+
\zeta (L_{-1}\otimes\bbbone)
+(\bbbone\otimes L_{0})$ to obtain
$$
\Delta_{(m',n')} \left\langle\psi_{(m',n')},
\pi\left(\psi_{(m,n)}\otimes\psi_{(1,2)}\right)\right\rangle
\hspace*{5.0cm}
\vspace*{-0.6cm}
$$
\begin{eqnarray}
\label{e1}
& = &
{\displaystyle
\left\langle\psi_{(m',n')},\pi\left(\zeta \left(L_{-1}\otimes\bbbone\right)
(\psi_{(m,n)}\otimes\psi_{(1,2)})\right)\right\rangle}
\nonumber \\
& &
{\displaystyle
+ (\Delta_{(m,n)} + \Delta_{(1,2)}) \left\langle\psi_{(m',n')},
\pi\left(\psi_{(m,n)}\otimes\psi_{(1,2)}\right)\right\rangle}
\end{eqnarray}
and similarly, by inserting $L_{0}^{2}$, we get
$$
\kappa (\kappa -1)  \left(\psi_{(m',n')},
\pi\left(\psi_{(m,n)}\otimes\psi_{(1,2)}\right)\right)
\hspace*{2.5cm}
$$
\begin{equation}
\label{e2}
\hspace*{2.0cm}
= \left\langle\psi_{(m',n')}, \pi \left( \left( \zeta^{2}
 (L_{-1}^{2}\otimes\bbbone)
\right)\left(\psi_{(m,n)}\otimes\psi_{(1,2)}\right)\right)
\right\rangle ,
\end{equation}
where $\kappa=\Delta_{(m',n')} - \Delta_{(m,n)} - \Delta_{(1,2)}$.
Finally,
\begin{eqnarray}
0 & = &
{\displaystyle
\left\langle\psi_{(m',n')},\pi\left(\Delta_{\zeta,0}\,(L_{0})\;
\Delta_{\zeta,0}\,(L_{-1})\;
\left(\psi_{(m,n)}\otimes\psi_{(1,2)}\right)\right)\right\rangle} \nonumber
\\
& = &
{\displaystyle \left(\Delta_{(m,n)} + \Delta_{(1,2)} +1 \right)
\left\langle\psi_{(m',n')},\pi\left(\Delta_{\zeta,0}\,(L_{-1})
\left(\psi_{(m,n)}\otimes\psi_{(1,2)}\right)\right)\right\rangle}
\nonumber \\
& & {\displaystyle +
\left\langle\psi_{(m',n')},\pi\left(\zeta (L_{-1}^{2}\otimes\bbbone)
\left(\psi_{(m,n)}\otimes\psi_{(1,2)}\right)\right)\right\rangle}
\nonumber \\
& & {\displaystyle +
\left\langle\psi_{(m',n')},\pi\left(\zeta(L_{-1}\otimes L_{-1})
\left(\psi_{(m,n)}\otimes\psi_{(1,2)}\right)\right)\right\rangle}
\nonumber \\
& = & {\displaystyle
\zeta^{-1} \kappa (\kappa -1)
\left\langle\psi_{(m',n')},\pi\left(
\left(\psi_{(m,n)}\otimes\psi_{(1,2)}\right)\right)\right\rangle}
\nonumber \\
& & {\displaystyle +
\left\langle\psi_{(m',n')},\pi\left(\zeta(L_{-1}\otimes L_{-1})
\left(\psi_{(m,n)}\otimes\psi_{(1,2)}\right)\right)\right\rangle,}
\end{eqnarray}
which implies
$$
\left\langle\psi_{(m',n')},\pi\left((L_{-1}\otimes L_{-1})
\left(\psi_{(m,n)}\otimes\psi_{(1,2)}\right)\right)\right\rangle
\hspace*{3.5cm} $$
\begin{equation}
\label{e3}
\hspace*{2.5cm}
= -\zeta^{-2} \kappa (\kappa -1) \left\langle\psi_{(m',n')},\pi\left(
\left(\psi_{(m,n)}\otimes\psi_{(1,2)}\right)\right)\right\rangle.
\end{equation}
Inserting (\ref{e1}), (\ref{e2}) and (\ref{e3}) into (\ref{null}) we obtain
\begin{equation}
\label{e4}
0=\zeta^{-2} \left( \kappa - \Delta_{(m,n)} +
\frac{3}{2(2\Delta_{(1,2)}+1)} \kappa (\kappa -1) \right)
\left\langle\psi_{(m',n')},\pi
\left(\psi_{(m,n)}\otimes\psi_{(1,2)}\right)\right\rangle.
\end{equation}
Hence, if the representation ${\cal H}_{(m',n')}$ is contained
in the tensor product of ${\cal H}_{(m,n)}$ and ${\cal H}_{(1,2)}$,
the bracket in (\ref{e4}) must vanish and
$\Delta_{(m',n')}$ must satisfy equation (5.21) of \cite{BPZ} ---
thus giving the well-known restrictions for the fusion rules in the case
where one of the fields is either $(1,2)$ or $(2,1)$
\begin{equation}
\label{f1}
\psi_{(1,2)} \otimes \psi_{(m,n)} = \left[ \psi_{(m,n-1)}\right]
\oplus \left[ \psi_{(m,n+1)}\right]
\end{equation}
and
\begin{equation}
\label{f2}
\psi_{(2,1)} \otimes \psi_{(m,n)} = \left[ \psi_{(m-1,n)}\right]
\oplus \left[ \psi_{(m+1,n)}\right].
\end{equation}
\smallskip

\noindent These equations imply in particular that the representation
$\psi_{(m,n)}$ is contained in
\begin{equation}
\left(\psi_{(2,1)}\right)^{\otimes (m-1)}\otimes \left(\psi_{(1,2)}\right)^
{\otimes (n-1)}.
\end{equation}
Thus, any representation contained in
$\psi_{(m_{1},n_{1})}\otimes \psi_{(m_{2},n_{2})}$ must also be
contained in
\begin{equation}
\left(\psi_{(2,1)}\right)^{\otimes (m_{1}-1)}\otimes
\left(\psi_{(1,2)}\right)^{\otimes (n_{1}-1)}\otimes
\psi_{(m_{2},n_{2})}
\end{equation}
and
\begin{equation}
\left(\psi_{(2,1)}\right)^{\otimes (m_{2}-1)}\otimes
\left(\psi_{(1,2)}\right)^{\otimes (n_{2}-1)}\otimes
\psi_{(m_{1},n_{1})},
\end{equation}
where we have (implicitly) used the associativity and the
symmetry of the tensor product.
Using (\ref{f1}) and (\ref{f2}) we can derive from these two
conditions the general restrictions for the fusion rules, namely
\begin{equation}
\psi_{(m_{1},n_{1})}\otimes \psi_{(m_{2},n_{2})}=
\sum_{k=|m_{1}-m_{2}|+1}^{m_{1}+m_{2}-1}\;
\sum_{l=|n_{1}-n_{2}|+1}^{n_{1}+n_{2}-1}\;
\left[ \psi_{(k,l)} \right],
\end{equation}
where the variable $k\; (l)$ attains only every other value.

To prove that these conditions are also sufficient, we
may apply the methods of \cite{BFIZ}, \cite{FF} or
\cite{K} to show, that the given irreducible representations
are actually contained in the tensor product.
In particular, our comultiplication is compatible
with \cite[(4.15)]{BFIZ}, which enables us to carry through the
proof given there.

We remark, that the map ${\cal F}$ of \cite{BFIZ}
has non-trivial kernel. This reflects the fact that
the true tensor product of conformal field
theory is actually the quotient of the vector space
tensor product, as described above.

\section{Conclusions}
\renewcommand{\theequation}{5.\arabic{equation}}
\setcounter{equation}{0}

In this paper we have shown that fusion in conformal field
theory can be understood as the ring-like tensor product
of modules of the symmetry algebra. We have proved that
the tensor product is associative and symmetric up to
equivalence. We also determined explicitly
the action of the symmetry algebra on these tensor products
and derived the fusion rules of the WZW- and the minimal models.

We have thus established a framework within which
determining the fusion rules is a well-posed
algebraic problem. We would like to point out, that
the above analysis can be generalized to a large class
of conformal field theories \cite{G}, thus providing
a powerful approach for determining the fusion rules
for many conformal field theories.
\smallskip

To get a better understanding of the r\^{o}le of the braiding
in conformal field theory we calculated the $R$-matrix
of the comultiplication. It turned out, that the $R$-matrix
is in fact triangular and that the braiding only occurs if
one tries to decompose the tensor product into
irreducible representations in a consistent way.
This seems to indicate, that the quantum group
relevant for conformal field theory does not
``sit inside'' the symmetry algebra.
On the contrary, we rather believe, that the quantum group
is connected to the fact, that in order to define
consistently associative (chiral) vertex operators
which only depend on the analytic parameter $z$,
one is forced to retain some information of the
action of the whole vertex operator on the
``antiholomorphic representation spaces''.
In fact, for the case of the WZW-models,
it turns out that it is sufficient to
retain the action on the lowest-energy states,
and these finite-dimensional vector spaces are
isomorphic to the representation spaces of the
corresponding quantum group $U_{q}(g)$.
This structure was already anticipated in \cite{MR},
however, a proper understanding is --- in our opinion ---
still missing.

\appendix

\section{Proof of comultiplication}

\renewcommand{\theequation}{A.\arabic{equation}}
\setcounter{equation}{0}

Let us first consider the case of the
Virasoro algebra. To prove the comulti\-pli\-ca\-tion-property
one has to treat the different cases --- corresponding
to the different values of $n$ in (\ref{vir1}), resp. (\ref{vir2})
--- separately.
Writing out both sides of equation (\ref{com}) the problem reduces
to proving (e.\ g.\  for the case $a=L_{n}, b=L_{k}$ and
$n,k \geq -1$ )
\begin{equation}
\sum_{l=\max{L-k,-1}}^{\min{L+1,n}}
\left( \begin{array}{c} n+1 \\ l+1 \end{array} \right)
\left( \begin{array}{c} k+1 \\ L+1-l \end{array} \right)
(2l - L) =
(n-k) \left( \begin{array}{c} n+k+1 \\ L+1 \end{array} \right).
\end{equation}
To prove this we observe, that
\begin{equation}
\sum_{m=-1}^{n}
\left( \begin{array}{c} n+1 \\ m+1 \end{array} \right)
z^{n-m} = (z+1)^{n+1}
\end{equation}
and
\begin{eqnarray}
{\displaystyle \sum_{m=-1}^{n}
\left( \begin{array}{c} n+1 \\ m+1 \end{array} \right)
z^{n-m} m} &  = & {\displaystyle \left. \frac{d}{d\varepsilon}
(z+\varepsilon)^{n+1} \right|_{\varepsilon=1} - (z+1)^{n+1}} \nonumber \\
& = & {\displaystyle (n+1) (z+1)^{n} - (z+1)^{n+1}}.
\end{eqnarray}
Hence we can calculate
\begin{equation}
\sum_{l_{1}=-1}^{n} \sum_{l_{2}=-1}^{k}
\left( \begin{array}{c} n+1 \\ l_{1}+1 \end{array} \right)
z^{n-l_{1}}
\left( \begin{array}{c} k+1 \\ l_{2}+1 \end{array} \right)
z^{k-l_{2}} (l_{1}-l_{2})=(n-k) (z+1)^{n+k+1}.
\end{equation}
On the other hand we can write the right-hand-side as
\begin{equation}
(n-k) \sum_{L=-1}^{n+k}
\left( \begin{array}{c} n+k+1 \\ L+1 \end{array} \right)
z^{n+k-L}
\end{equation}
and comparing the coefficients of the Cauchy product of the two sums
in (A.4) with (A.5) we derive the desired formula (A.1).
\smallskip

The other cases can be dealt with similarly, once one observes that
\begin{eqnarray}
{\displaystyle \sum_{m=-1}^{\infty}
\left( \begin{array}{c} n+m-1 \\ m+1 \end{array} \right)
(-1)^{m+1} z^{-(n+m)}} & = &
{\displaystyle \frac{(-1)^{n-1}}{(n-2)!}\;
\frac{d^{n-2}}{dz^{n-2}}
\sum_{m=-1}^{\infty} (-z)^{-(m+2)} } \nonumber \\
& = & {\displaystyle (z+1)^{1-n}}
\end{eqnarray}
and
\begin{eqnarray}
{\displaystyle \sum_{m=-1}^{\infty}
\left( \begin{array}{c} n+m-1 \\ m+1 \end{array} \right)
(-1)^{m+1} z^{-(n+m)} m} & = &
{\displaystyle \left( -z \frac{d}{dz} -n \right) (z+1)^{1-n}} \nonumber \\
& = & {\displaystyle - (z+n) (z+1)^{-n}}.
\end{eqnarray}
\medskip

The proof in the case of the Kac-Moody algebra goes along
similar lines. We would just like to point out that
\begin{eqnarray}
{\displaystyle \sum_{m=0}^{\infty}
\left( \begin{array}{c} n+m-1 \\ m \end{array} \right)
(-1)^{m} z^{-(n+m)}} & = &
{\displaystyle \frac{(-1)^{n}}{(n-1)!}
\frac{d^{n-1}}{dz^{n-1}} \left(\frac{1}{1+z^{-1}} -1 \right)} \nonumber \\
& = & {\displaystyle (z+1)^{-n}}
\end{eqnarray}
and
\begin{eqnarray}
{\displaystyle \sum_{m=n}^{\infty}
\left( \begin{array}{c} m-1 \\ n-1 \end{array} \right)
(-z)^{m-n}} & = &
{\displaystyle \frac{(-1)^{n-1}}{(n-1)!} \frac{d^{n-1}}{dz^{n-1}}
\sum_{m=1}^{\infty} (-z)^{m-1}} \nonumber \\
& = & {\displaystyle (z+1)^{-n}.}
\end{eqnarray}

\section{Coassociativity}
\renewcommand{\theequation}{B.\arabic{equation}}
\setcounter{equation}{0}

We give the proof of the coassociativity formula (\ref{coass}) for
the case of the Virasoro algebra, when we apply both
sides to $L_{-m}$ with $m >2$. There are three identities to check,
corresponding to terms of the form
$L_{k}\otimes\bbbone\otimes\bbbone,\;\; \bbbone\otimes L_{k}\otimes
\bbbone$ and $\bbbone\otimes\bbbone\otimes L_{k}$, respectively.
The left-hand-side corresponding to the first term is
\begin{equation}
\sum_{l=-1}^{\infty}
\left( \begin{array}{c} l+m-1 \\ m-2  \end{array} \right)
(-1)^{l+1} \zeta_{1}^{-(m+l)}\;
\sum_{k=-1}^{l}
\left( \begin{array}{c} l+1 \\ k+1  \end{array} \right)
(\zeta_{2} - \zeta_{1})^{l-k}
\left( L_{k}\otimes\bbbone\otimes\bbbone \right),
\end{equation}
which can be resummed obtaining
\begin{equation}
\sum_{k=-1}^{\infty}
\left(\;
\sum_{l=k}^{\infty}
\left( \begin{array}{c} l+m-1 \\ m-2  \end{array} \right)
(-1)^{l+1}
\left( \begin{array}{c} l+1 \\ k+1  \end{array} \right)
(\zeta_{2} - \zeta_{1})^{l-k} \zeta_{1}^{-(m+l)} \right)
\left( L_{k}\otimes\bbbone\otimes\bbbone \right).
\end{equation}
The sum in the brackets can be calculated to be
$$
\frac{(k+m-1)! \; (-1)^{(k+1)}}{(k+1)! \; (m-2)!}
\sum_{l=k}^{\infty} \frac{(l+m-1)!}{(m+k-1)! \; (l-k)!} \; (-1)^{l-k} \;
\zeta_{1}^{-(m+l)}\; (\zeta_{2} - \zeta_{1})^{l-k}  \hspace*{0.5cm}
$$
\begin{eqnarray}
& = &
{\displaystyle
\left( \begin{array}{c} k+m-1 \\ k+1  \end{array} \right) (-1)^{k+1}
\frac{(-1)^{m+k-1}}{(m+k-1)!} \;
\frac{d^{m+k-1}}{dz_{1}^{m+k-1}} \left.
\sum_{l=0}^{\infty} z_{1}^{-(l+1)} (\zeta_{1} - \zeta_{2})^{l}
\right|_{z_{1}=\zeta_{1}}} \nonumber \\
& = & {\displaystyle
\left( \begin{array}{c} k+m-1 \\ k+1  \end{array} \right) (-1)^{k+1}
\zeta_{2}^{-(m+k)}},
\end{eqnarray}
thus agreeing with the right-hand-side of (\ref{coass}) corresponding to
the first term.
The other terms can be dealt with similarly.
\bigskip

\noindent {\bf Acknowledgements}
\smallskip

It is a pleasure to thank my PhD supervisor Peter Goddard for much
advice and encouragement. I also want to thank
Richard Borcherds for helpful remarks and
M. Chu, H. Kausch, A. Kent, S. Majid, G. Segal and G. Watts for useful
discussions.

I am grateful to Pembroke College, Cambridge,
for a research studentship and to the Studienstiftung des deutschen
Volkes for financial support.

\end{document}